# Forecasting model based on information-granulated GA-SVR and ARIMA for producer price index

Xiangyan Tang [1], Liang Wang[2,*], Jieren Cheng[1,3], Jing Chen[1], Victor S. Sheng[4]

**Abstract:** The accuracy of predicting the Producer Price Index (PPI) plays an indispensable role in government economic work. However, it is difficult to forecast the PPI. In our research, we first propose an unprecedented hybrid model based on fuzzy information granulation that integrates the GA-SVR and ARIMA (Autoregressive Integrated Moving Average Model) models. The fuzzy-information-granulation-based GA-SVR-ARIMA hybrid model is intended to deal with the problem of imprecision in PPI estimation. The proposed model adopts the fuzzy information-granulation algorithm to pre-classification-process monthly training samples of the PPI, and produced three different sequences of fuzzy information granules, whose Support Vector Regression (SVR) machine forecast models were separately established for their Genetic Algorithm (GA) optimization parameters. Finally, the residual errors of the GA-SVR model were rectified through ARIMA modeling, and the PPI estimate was reached. Research shows that the PPI value predicted by this hybrid model is more accurate than that predicted by other models, including ARIMA, GRNN, and GA-SVR, following several comparative experiments. Research also indicates the precision and validation of the PPI prediction of the hybrid model and demonstrates that the model has consistent ability to leverage the forecasting advantage of GA-SVR in non-linear space and of ARIMA in linear space.

**Keywords:** Data analysis, producer price index, fuzzy information granulation, ARIMA model, support vector model.

## 1 Introduction

The Producer Price Index (PPI) is a family of indexes used to measure the average change over time in selling prices received by domestic producers of goods and services. PPIs measure price changes from the perspective of the seller. This contrasts with other measures, such as the Consumer Price Index (CPI), that measure price changes from the purchaser's perspective. Sellers' and purchasers' prices may differ due to government subsidies, sales and excise taxes, and distribution costs. The PPI plays an indispensable role in fields requiring contract adjustment, indicators of overall price movement at the producer level, deflators of other economic series, measure of price movement for particular industries and products, comparison of input and output costs, comparison of industry-based price data to other industry-oriented economic time series, and forecasting. The industrial producer price index includes the "Producer Price Index for Industrial Producers" and "Purchase Price Index for Industrial Producers," and the former is generally called the Producer Price Index (referred to as the PPI in this paper). PPI, a relative number that reflects the trend as well as variation of ex-factory prices of all industrial products within a certain period of time, contains various products sold by industrial enterprises to all enterprises except itself and the products sold directly to residents for consumption, which makes it possible to determine the impact of ex-factory price changes on gross industrial output value and added value. Since the PPI reflects the trend and the variation of the ex-factory prices of the industrial products for the first sale, if it is higher than the expected figure, there exists the risk of inflation; when lower than expected, it indicates a risk of deflation. As a result, the PPI is extremely sensitive to the national economy, which is not only used in business negotiations and economic analysis but is also an important price variable for International Monetary Fund (IMF) to compare price levels among different countries or economies. [Agdas, Pesantes-Tavares, Agdas et al. (2011)] used the PPI as an important price-adjustment index for solving the problem of price fluctuations of certain commodities, such as oil, steel, and cement. [Norman (2004)] highlighted the significance of the PPI in services in Europe. Unfortunately, due to a variety of uncertain factors, predicting PPI accurately is a difficult problem, yet to be adequately resolved it requires exploring a new way of scientifically measuring the variation range and the trend of the PPI, which undoubtedly is of forward-looking significance in grasping the law of price fluctuations of industrial products, estimating future price fluctuations, and avoiding market risks to develop policies.

[1] College of Information Science & Technology, Hainan University, Haikou 570228, China.

[2] School of Economic and Management, Hainan University, Haikou 570228, China.

[3] State key laboratory of Marine Resource Utilization in South China Sea, Haikou 570228, China.

[4] Department of Computer Science, University of Central Arkansas, Conway, AR 72035, USA.

[*] Corresponding Author: Liang Wang. Email: cjr@hainu.edu.cn.

## 2 Literature review

Parametric models are widely applied to predict PPI and other economic indicators in current research, such as [Wang (2016)] have used time-series co-integration regression and [Hua, Wu, Li et al. (2016); Cheng, Xu, Tang et al. (2018)] have used the ARIMA model for short-term forecasting. Some scholars have focused on the non-linear and chaotic timing characteristics of PPI, e.g., [Wang, Gao, Liu et al. (2017)] used the normality test, correlation dimension test, and largest Lyapunov method to examine the time series of Chinese PPIs from 1996 to 2016, uncovering a strong chaotic property and complicated evolution law. However, the traditional method of parameter prediction fails to capture the real law of index changes. In contrast, the scientific nature of non-parametric methods has gradually become prominent. For example, [Su and Yuan (2013)] introduced an improved SVR model with a kernel principal component genetic algorithm to predict the stock index, proving that the non-parametric method has a great edge in predicting chaotic time series. [Wang, Li and Li (2009)] discovered the non-linear and chaotic characteristics of China's CPI via a BP (Back Propagation) neural network with momentum. [Yang and Li (2011)] compared the different effects of option pricing among the non-parametric method SVR, binary tree, Monte Carlo model, and other traditional parametric models, reaching the conclusion that the traditional parametric models cannot have good commend of the trends and non-linear characteristics of the actual option price. [Tay and Cao (2001); Cao and Tay (2001)] compared the ability of a BP neural network and SVR model in predicting financial time series, and the latter model turns out to be superior to the former. [Zhang, Liu, Jin et al. (2017)] applied principal component regression (PCR), partial least-squares regression (PLSR), and SVR models to explore an effective method for predicting phenolics contents, and the results prove that SVR behaves globally better than PLSR and PCR. [Zhu, Lian, Liu et al. (2017)] integrated SVR into the hybrid model to effectively forecast air pollution. [Čeperić, Žiković S and Čeperić (2017)] presented the results of short-term forecasting of Henry Hub spot natural gas prices based on the performance of classical time-series models and machine-learning methods, which specifically are neural networks (NNs) and strategic seasonality-adjusted support vector regression (SSA-SVR) machines. [Das and Padhy (2017)] introduced a hybrid model called USELM-SVR, which can be useful as an alternative model for prediction tasks when more accurate predictions are required. [Hua, Wu, Li et al. (2016)] developed a model to predict the silicon content using SVR combined with clustering algorithms, which serves better for practical production. [Sermpinis, Stasinakis and Hassanniakalager (2017)] introduced a reverse adaptive Krill Herd–locally weighted SVR (RKI-ILSVR) model that outperforms its counterparts in terms of statistical accuracy and trading efficiency.

Concerning the positive effect of the parameter model in predicting linear trends, scholars have improved the non-parametric prediction model from the perspective of parameter optimization, and the achievements can be summarized in two aspects. The first is the improvement in particle swarm optimization (PSO), the genetic algorithm (GA), and other non-parametric models. For example, [Chen, Liang, Lu et al. (2014)] put forward the GA-based SVR model, the APSO-based SVR model, and the optimal forecasting model based on seasonal SVR and PSO, all of which make the best of non-linear prediction, parameter optimization, SVR's dealing with small samples and nonlinear prediction characteristics, and APSO optimization on SVR parameters, ultimately forecasting the daily flow of tourist attractions. The second is the establishment of a hybrid model that consists of the non-parametric model and the traditional parametric model. [Xiong (2011)] integrated ARIMA with NNs to forecast the RMB exchange rate in terms of the linear and non-linear space prediction of the two models. [Liang, Ma, Chen et al. (2015)] established the SVR-ARIMA hybrid model to make a comparison with SVR or the ARIMA model, and the final results demonstrate that the hybrid model is more robust and accurate in prediction.

Regarding data pre-processing, a few scholars have made breakthroughs in further study. [Su and Fu (2013)] employed the principal component analysis method to reduce the dimension of original data. [Lu, Zhao and Bi (2015)] applied fuzzy information granulation to depose CPI time-series data, finally obtaining good results on the prediction of the processed data. In particular, in the field of traffic-flow forecasting, [Sun, Shao, Ji et al. (2014)] pre-processed data through information granulation and proposed the ARIMA-SVR hybrid model, leading to a better prediction, which further reinforces the information-granulation method of pre-processing data scientifically.

As noted above, it has been found that the previous studies have a deficiency, namely overusing the fuzzy information-granulation method, parameter optimization support vector regression model, and ARIMA model alone; however, the improvement of SVR parameter optimization and the great advantage of the hybrid model inspire us to great extent. Hence, the adoption of predictive algorithms considering the relevance of many model methods is expected to be the main emphasis. Overall, this paper combines GA-SVR with ARIMA based on fuzzy information granulation and tries to establish a GA-SVR model, and then corrects the predicted results using the ARIMA model. Finally, a GA-SVR-ARIMA hybrid model based on fuzzy information granulation is completed, attempting to fully leverage the advantages of the GA-SVR in non-linear space prediction and the ARIMA in linear space prediction to achieve the ideal prediction of PPI.

## 3 Establishment of information-granulated GA-SVR-ARIMA hybrid model

Here, we divide the establishment of the Information-Granulated GA-SVR-ARIMA Hybrid Model into the following processes:

### *3.1 Information-granulated GA-SVR-ARIMA hybrid model process*

1) Obtain monthly PPI data and then input them.
2) Take one-quarter (3 months) as the information-granulation window, and then output the three sequences after the fuzzy information granulation is disposed on the PPI data.
3) Use the GA and MSE fitness function (objective function) to carry out fivefold cross-validation on the above sequences to seek the optimal penalty parameters C, the insensitive loss function $\varepsilon$, and the parameter $\gamma$ in the RBF kernel function corresponding to the SVR machine model, respectively.
4) Specify the training set and test set and establish the GA-SVR model for training and prediction according to the optimal penalty parameters C, the insensitive loss function $\varepsilon$, and the parameter $\gamma$ in the RBF kernel function of the above three fuzzy information-granule sequences.
5) Output the residuals of the trained GA-SVR model, and then use the ARIMA model to train the GA-SVR residuals in an effort to correct the prediction results.
6) Establish the GA-SVR-ARIMA hybrid model and output the prediction results of the three fuzzy information-granule sequences, which present the trend and variation of the PPI.

The model flowchart is presented in Fig.1.

In general, the fuzzy information-granulated method enables us to seek out the optimal solution, reduce the dimension, and improve the convergence speed of the fitting process through collaborative information sharing in the group. SVR has excellent generalization ability and advantages in handling small samples, nonlinearity, and high-dimensional space, so that problems such as dimensionality disaster and local extrema can be effectively alleviated. The GA distinguishes itself by high computing efficiency and finding the global optimal solution with probabilistic search technology. The ARIMA model is more suitable for linear time-series prediction. Therefore, the paper builds the GA-SVR-ARIMA model based on fuzzy information granulation that is well integrated with the strengths of the aforementioned methods to resolve the problem of PPI prediction.

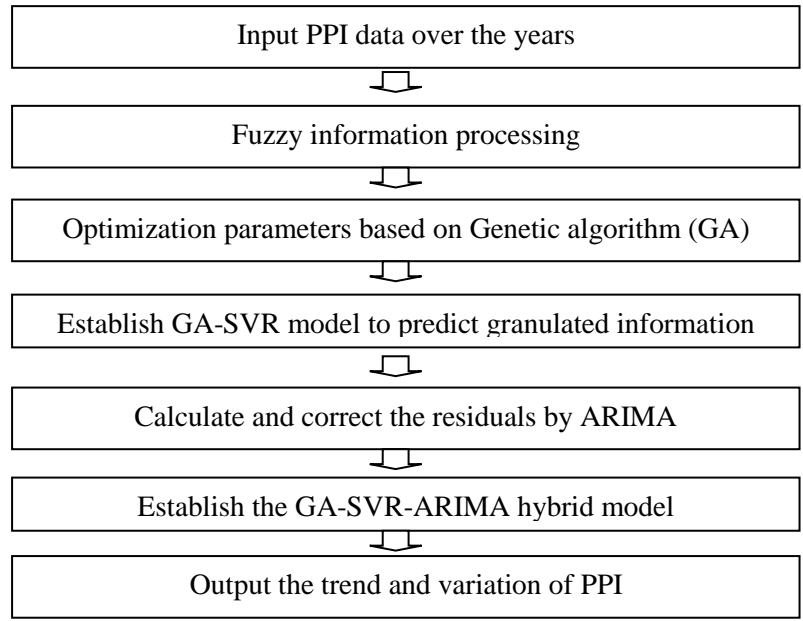

**Figure 1:** Algorithm flowchart of Information Granulated GA- SVR-ARIMA hybrid model

*3.2 Pre-classification of fuzzy information granulation*

An information granule is a clump of objects drawn together by indistinguishability, similarity, proximity, or functionality, and its density depends entirely on the boundary definition of the information granule. The concept of the information granule was first proposed and discussed on the basis of fuzzy set theory by [Zadeh (1979)] who supported dividing amounts of complex information into several simple blocks and viewing each block as an information granule.

There are three types of information granulation: fuzzy set theory, rough set theory, and quotient space theory. In particular, fuzzy information granules are information granules represented by fuzzy sets. The process of fuzzy information granulation in a time series comprises two steps: window division and fuzzification. Specifically, the window partition is used to divide the time series into several sub-sequences as the operation window, while the fuzzification is used to turn the divided windows into fuzzy sets in order to replace the data in the original window. In short, the combination of these two generalized models is fuzzy information granulation, which is referred to as f granulation. The kernel of granulation is to set up a reasonable fuzzy set for the given window so that it can take the place of the data in the original window. In this research, we make a given time series fuzzy and establish a fuzzy information particle P on X—a fuzzy concept G that can reasonably describes X (a fuzzy set regarding X as the Universe):

$$G \text{ is defined as } P \triangleq X \tag{1}$$

G, in essence, can determine the course of a function P=A(x), and A is the membership function of the fuzzy concept G. In this paper, the triangular fuzzy particles are chosen to be studied in the method of W.pedrycz granulation. The membership function of triangular fuzzy particles can be expressed as

$$A(x, a, m, b) = \begin{cases} 0, x < a; \\ \dfrac{x-a}{m-a}, a \leq x \leq m; \\ \dfrac{b-x}{b-m}, m < x \leq b; \\ 0, x > b. \end{cases} \tag{2}$$

where X is the time series, and a, m, and b are parameters. Regarding the single fuzzy particle, the three parameters represent the minimum, the average and the maximum change, respectively, of the corresponding raw data.

### *3.3 Parameter optimization based on genetic algorithm*

The GA was first proposed by [Holland (1998)] and is based on Darwin's theory of evolution. In this paper, we adopt the GA to search for the best penalty parameter C, the insensitive loss function $\varepsilon$, and the parameter $\gamma$ of the RBF kernel function in the SVR model, and then construct the GA-SVR model to predict the fuzzy information-granule sequences of PPI monthly data. The specific steps are as follows:

1) Choose a coding strategy including population size, selection, crossover, and mutation methods, and genetic parameters such as crossover probability $P_c$ and mutation probability $P_m$. Since the individual is assessed by its fitness to determine the genetic chance in the GA, the paper select the fitness function mean-square error (MSE) with the evolutionary algebra of 200 generations and the population of 20. MSE, a measure of the mean-square error generated from a subset of cross-validation (CV) mechanisms, can effectively measure the chromosome quality in regression prediction and avoid or mitigate over-fitting after cross-validation. In this paper, fivefold CV is carried out, and the fitness function formula is as follows:

$$MSE = \frac{1}{n} \sum_{i=1}^{n} (y_p - y_p^*)^2 \tag{3}$$

where $y_p$ is the observed value, $y_p^*$ is the predictive value, and n represents the training sample set of the fuzzy information particles. In addition, the individual effect is better or the probability of being selected is higher in the case of the smaller fitness value.

2) Code according to the characteristic subset of each chromosome, and then randomly generate initialization population P. The selection of coding scheme usually depends on the nature of the problem remaining to be solved. Moreover, the common encoding schemes are binary coding and real coding, and the former is widely used.
3) Calculate the fitness value of each individual in the population according to the fitness function. Then, utilize selection, cross-over, and mutation operators for genetic manipulation to form the next generation of groups.
4) Determine whether the fitness value meets the predetermined criterion; if not, return to the previous step, or return to step (2), and continue to execute the optimization algorithm to reach termination. Eventually, take the individual with the smallest fitness in the evolution process as optimal.

The SVR model performs well in classification or regression, but the generalization relies on parameter setting to a great extent. For a given dataset, it is essential to find the optimal parameter. However, in practical applications, the problem of parameter selection has yet to be properly solved, and it is mainly selected by experiments or a time-consuming grid-search method.

The GA, a robust search algorithm for optimizing a complex system, possesses special superiority compared with other intelligent algorithms. Owing to the natural selection of the fittest and the simple genetic operation, the GA is not constrained by restrictions such as the search space while calculating, and does not need any auxiliary information, which allows it to find the global optimal solution easily. However, the GA can search with multiple search points at the same time by the method of population organization, which is especially suitable for parallel processing of large-scale data, leading to higher computation efficiency.

### 3.4 Training and prediction of support vector machine

A support vector machine is a neural network model for studying small samples and small-probability events that was proposed by [Vapnik and Vladimir (2002)] in the 1990s, and later introduced into the field of regression prediction, in which it was called SVR. SVR implements the function of regression estimation based on the principle of structural risk minimization, which is estimated by the insensitive loss function $\varepsilon$. Furthermore, SVR makes use of the risk function that is well integrated with the minimization principle of empirical error and structural risk. In this paper, the construction principle of the nonlinear $\varepsilon$-SVR model is expressed as follows:

Given a dataset $G = \{(x_i, y_i)\}_i^n$, $x_i$ is the input eigenvector, $y_i$ is the target value, and n is the sample size of time-series data. The basic idea of non-linear SVR is to map data x to a high-dimensional feature space through a nonlinear mapping and to conduct the linear regression in this space:

$$f(x) = \omega^T \cdot \Phi(x) + b \tag{4}$$

$$\Phi: R^n \to F, \quad \omega \in F \tag{5}$$

In Eq. 4 and Eq. 5, b is the threshold value and $\Phi$ is the high-dimensional feature space, which is a non-linear mapping of inputting X. The problem of optimization in estimating $\omega$ and b can be calculated by the minimum of the following equation:

$$\frac{1}{2}\|w\|^2 + C\sum_{i=1}^{i}(\xi_i + \xi_i^*)$$

$$s.t \begin{cases} y_i - w \cdot \varphi(x) - b \leq \varepsilon + \xi_i \\ w \cdot \varphi(x) + b - y_i \leq \varepsilon + \xi_i^* \\ \xi_i, \xi_i^* \geq 0, i = 1,....,n \end{cases} \quad (6)$$

where C is the penalty parameter, $\xi_i, \xi_i^*$ are the relaxation variables, $\varepsilon$ is their sensitive loss function, and the introduction of X improves the robustness of the estimation. In the empirical study, the parameters C and $\varepsilon$ should be selected by the researchers themselves. Moreover, the duality theory is generally adopted to transform the above problem into a convex quadratic programming problem. Then, employing Lagrange's transformation on Eq. 6 gives

$$L = \frac{1}{2}\|\omega\|^2 + C\sum_{i=1}^{i}(\xi_i + \xi_i^*) - \sum_{i=1}^{i}\lambda_i(\varepsilon + \xi_i - y_i + (\omega, x_i) + b)$$
$$-\sum_{i=1}^{i}\lambda_i^*(\varepsilon + \xi_i^* + y_i - (\omega, x_i) - b) - \sum_{i=1}^{i}(\eta_i\xi_i + \eta_i^*\xi_i^*) \quad (7)$$

In Eq. 7, $\lambda_i、\lambda_i^*、\eta_i、\eta_i^* \geq 0, \ i = 1,...,n$, and the partial derivative of the Lagrange function with respect to the variables $\omega$, b, $\xi_i$, and $\xi_i^*$ is zero. After importing Lagrange formulas and optimization restrictions, the decision function of Eq. 7 is transformed into

$$f(x) = \sum_{i=1}^{l}(\lambda_i - \lambda_i^*)k(x, x_i) + b \quad (8)$$

where $k(x, x_i)$ is the kernel function of the SVM. When dealing with non-linear problems, the low-dimensional non-linear original data can be mapped to the high-dimensional feature space through the kernel function, where it has access to the linear treatment. The commonly used kernel functions are mainly the linear kernel function, the polynomial kernel function, and the Gaussian radial basis function (RBF). From previous experience, the RBF is superior in effect in the case of a lack of prior knowledge of sample data, so the RBF is accepted as the kernel function, which is expressed as

$$k(x, x_i) = \exp(-\gamma \|x - x_i\|^2), \gamma > 0 \quad (9)$$

In Eq. 9, $\gamma$ is the kernel parameter, the selection of which has a significant impact on the kernel function. In detail, the overfitting is caused by the large setting, while the weak generalization can be attributed to the small setting.

### *3.5 ARIMA model error correction method*

The autoregressive integrated moving average model (ARIMA), a time-series modeling method proposed by [Box and Jenkins (1973)] is virtually an extension of the autoregressive moving average model (ARMA). Accordingly, the paper applies the ARIMA model to the prediction residual of the three fuzzy information particles in PPI monthly time-series data to correct the deviation of GA-SVR, and the ARIMA model can be expressed as

$$\Delta^d y_t = \theta_0 + \sum_{i=1}^{p} \phi_i \Delta^d y_{t-i} + u_t + \sum_{j=1}^{q} \theta_j u_{t-j} \tag{10}$$

where $y_t$ represents the prediction residual of fuzzy information particles in PPI monthly data, and $\Delta^d y_t$ the sequence after d differential conversion of $y_t$. Regarding $u_t$, it is the random error at time t that represents the white-noise sequence subject to the normal distribution with a mean of 0 and constant variance $\sigma^2$. Furthermore, $\phi_i$ (i=1,2,..,p) and $\theta_j$ (j=1,2,…,q) are the parameters to be estimated; that is, p and q are the orders of the model. In short, the above model is denoted ARIMA (p, d, q), which, in essence, is a linear model, making it more suitable for linear modeling.

The flowchart of ARIMA modeling and prediction in this paper is shown in Fig. 2.

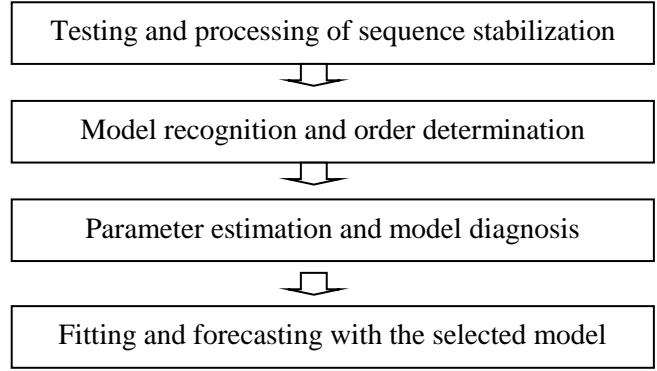

**Figure 2:** Algorithm flowchart of ARIMA model

Using Eviews 8.0, we build the final GA-SVR-ARIMA hybrid model by using the ARIMA model for the GA-SVR prediction residuals of three parameters—Min, Mean, and Max—in the training set. In addition, we also build an ARIMA model for the original PPI sequence for later comparative experiments. Now we firstly need to do three necessary steps before the final step ('Fitting and forecasting with the selected model').

*3.5.1 Residual visualization and ADF stationarity test*

In Fig. 3, the GA-SVR residuals of the above parameters all fluctuate around 0 value, and there is no obvious trend term nor is there a constant term. Under the circumstances, we examined the three series through the ADF stationarity test without a trend term and constant term.

In Fig. 4, no obvious trend term exists in the original PPI training sequence, but the constant item is significant, and hence the ADF stationarity test is used on the original PPI sequence with only a constant item, and the results are shown in Tab. 1.

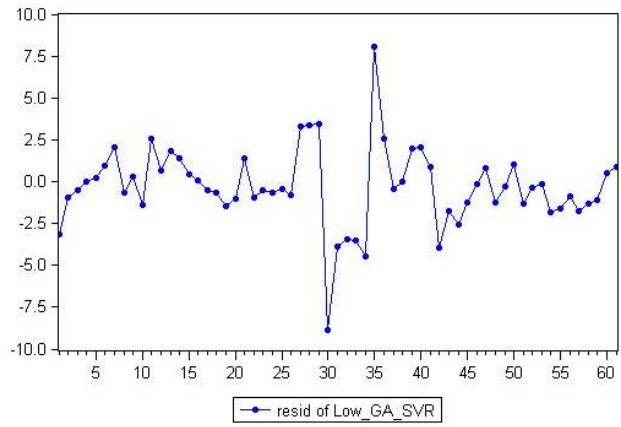

(a) Min model residual

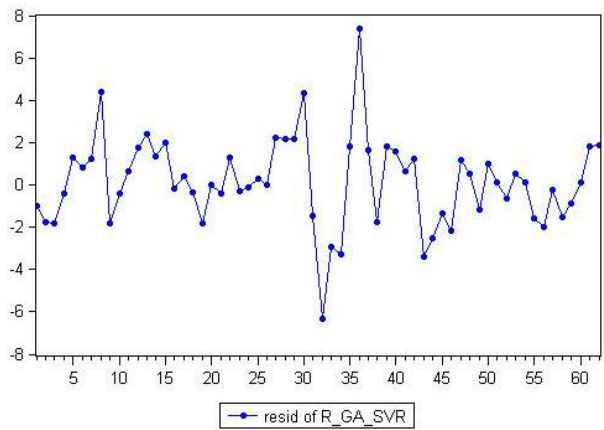

(b) Mean model residual

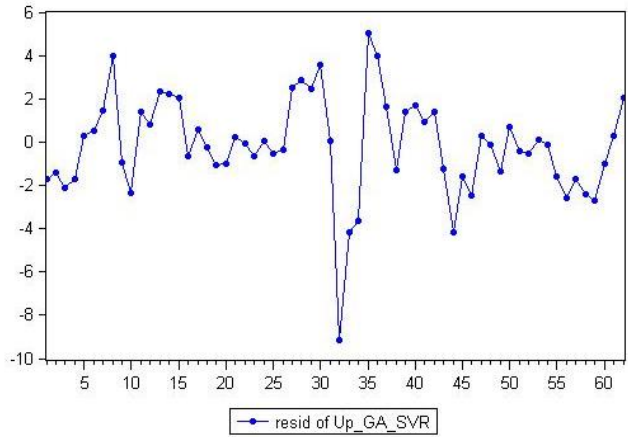

(c) Max model residual

**Figure 3:** GA-SVR model residual

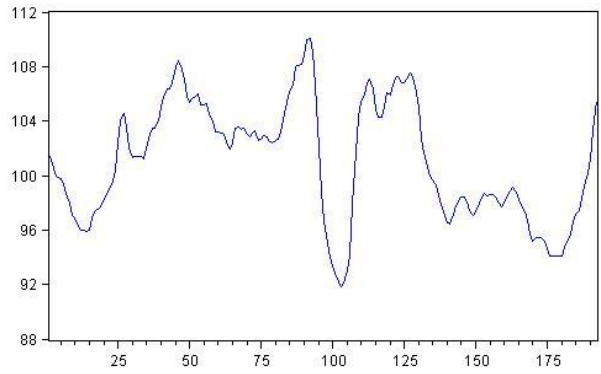

**Figure 4:** PPI sequence diagram

**Table 1:** ADF stationarity test results

| Residual series | | t-Statistic | Prob.* |
|---|---|---|---|
| Min | Augmented Dickey-Fuller test statistic | −6.112856 | 0.0000 |
| Mean | Augmented Dickey-Fuller test statistic | −5.795126 | 0.0000 |
| Max | Augmented Dickey-Fuller test statistic | −5.474432 | 0.0000 |
| PPI original sequence | Augmented Dickey-Fuller test statistic | −3.172854 | 0.0232 |

*MacKinnon (1996) one-sided p-values.

From Tab. 1, it is noted that the original residual series are both at 1% significance level, which can be recognized as a stationary series; hence, it is not necessary to differentiate them, or, in other words, the difference order is zero.

The Prob<0.05 of the ADF stationarity test of the original PPI sequence states that it can be regarded as a stationary series, so it is unnecessary to differentiate them; in other words, the difference order is zero.

*3.5.2 Model recognition and order determination*

We preliminarily reach a decision from Fig. 5 that the orders of the GA-SVR model residual sequence of Min are described as follows: p=1–5 and q=1–5. Accordingly, the orders of the GA-SVR model residual sequence of Mean are p=1–-4 and q=1–4, and the orders of the GA-SVR model residual sequence of Mean are p=1–4 and q=1–3. Next, we determine the Min model to be ARIMA (1, 0, 5), the Mean model ARIMA (1, 0, 2), and the Max model ARIMA (1, 0, 2) based on the analysis of the AIC (Akaike Information Criterion) minimum criterion.

(a) Min

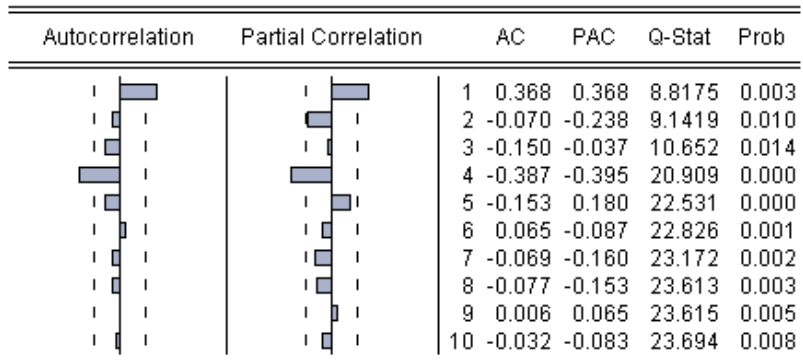

(b) Mean

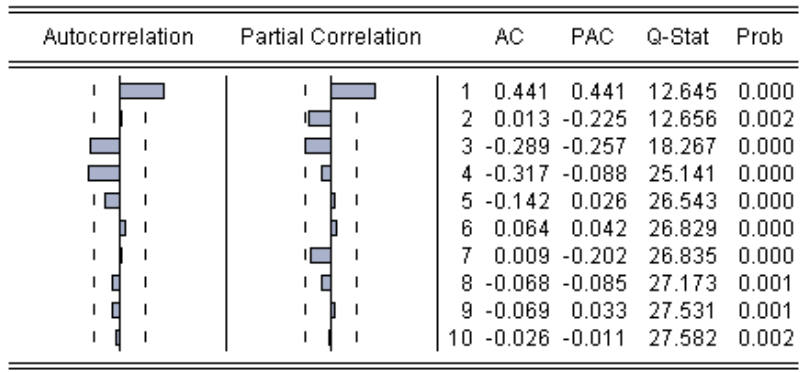

(c) Max

**Figure 5:** Autocorrelation and partial autocorrelation of residual sequence

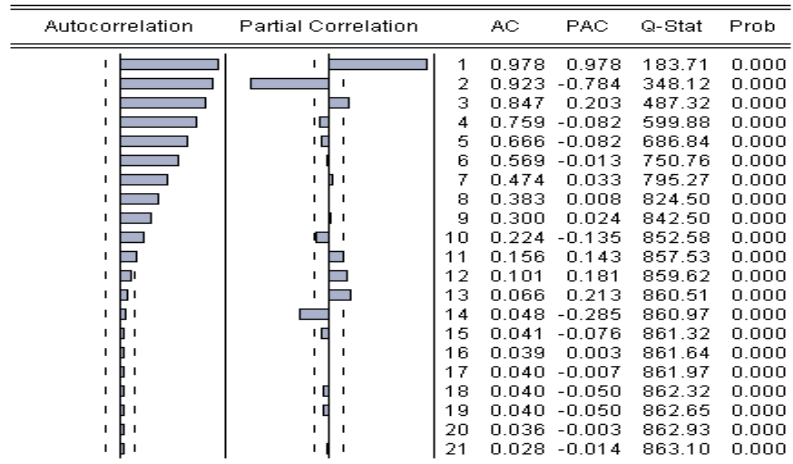

**Figure 6:** Autocorrelation and partial autocorrelation of original PPI sequences

**Table 2:** Parameter estimation and test results

| Model | Variable | Coefficient | Std. Error | t-Statistic | Prob. |
|---|---|---|---|---|---|
| Min: ARIMA (1, 0, 5) | AR (1) | 0.274014 | 0.12586 | 2.18860 | 0.032 |
|  | MA (5) | −0.471181 | 0.11943 | −3.95122 | 0.000 |
| Mean: ARIMA (1, 0, 2) | AR (1) | 0.542553 | 0.12031 | 4.446027 | 0.000 |
|  | AR (2) | −0.596258 | 0.11105 | −5.13596 | 0.000 |
| Max: ARIMA (1, 0, 2) | AR (1) | 1.144528 | 0.22574 | 5.073838 | 0.000 |
|  | AR (2) | −0.530452 | 0.12028 | −4.40943 | 0.000 |
|  | MA (1) | −0.656085 | 0.24016 | −2.66843 | 0.009 |
| Original PPI sequence: ARIMA (9, 0, 3) | AR (1) | 1.971104 | 0.07687 | 25.64104 | 0.000 |
|  | AR (2) | −1.232291 | 0.16985 | −7.25506 | 0.000 |
|  | AR (3) | −0.610441 | 0.19722 | −3.09512 | 0.002 |
|  | AR (4) | 1.751455 | 0.20891 | 8.383448 | 0.000 |
|  | AR (5) | −1.067294 | 0.23147 | −4.61081 | 0.000 |
|  | AR (6) | −0.005820 | 0.20421 | −0.02849 | 0.977 |
|  | AR (7) | 0.325880 | 0.19760 | 1.649172 | 0.101 |
|  | AR (8) | −0.129274 | 0.17091 | −0.75636 | 0.450 |
|  | AR (9) | −0.003238 | 0.07713 | −0.04198 | 0.966 |
|  | MA (3) | 0.930459 | 0.03098 | 30.03087 | 0.000 |

Eviews 8.0 is then used to draw the autocorrelation and partial correlation diagrams of the original PPI training sequence to judge the orders of the experimental model. As shown in Fig. 6, the orders of the original PPI sequences are initially determined as follows: p=9 and q=1–3. Next, we determine the model to be ARIMA (9, 0, 3) based on the analysis of the AIC minimum criterion.

*3.5.3 Parameter estimation and model diagnosis*

In Eviews 8.0, the ordinary least square method (OLS) is used to estimate the parameters of the three models and of ARIMA (9,0,3) about the original PPI sequence, and the results are summarized in Tab. 2. Regarding the test results of model parameters, it is clear that the AR (1) of the ARIMA (1,0,5) is significant at the level of the 5% confidence interval, and the other parameters are all significant at 1% confidence level. However, the conclusion is untenable in the model Min.

To confirm the reasonability, it is indispensable to test the white noise of the prediction residuals of the ARIMA model. Therefore, we draw the autocorrelation and partial autocorrelation functions of the fitted residuals of the three ARIMA models using Eviews 8.0, and the white-noise test is carried out with the Q-statistic in the BOX-Ljung test to verify the adaptability of the established model. The results are shown in Fig. 7.

From Fig. 7, the Prob-values of the Q-statistic for each order are all greater than 0.1. At the 10% significance level, accepting the original fake "This sequence is a white noise sequence" means that the residual sequence estimated by the ARIMA model that we built is a purely random sequence, so it is of no significance to model and search for other ARIMA models any longer, which is consistent with the rationality of the model built.

To prove the reasonability, it is necessary to test the white noise of the prediction residuals of the ARIMA (9,3,0) model. Therefore, we draw the autocorrelation and partial autocorrelation functions of the fitted residuals of the ARIMA (9,3,0) model using Eviews 8.0, and the white-noise test is carried out to verify the adaptability of the established model. The results are shown in Fig. 8.

From Fig. 8, the P-values of the Q-statistic for each order are all greater than 0.1. At the 10% significance level, accepting the original fake "This sequence is a white noise sequence" means that the residual sequence estimated by the ARIMA model is a purely random sequence, so it is of no statistical significance, which, in turn, verifies the rationality of the model.

## 4 Experiment

### 4.1 Experimental datasets and evaluation criterion

The survey on industrial producer prices covers the prices of more than 20,000 industrial products in 1638 basic categories. In this paper, we take the PPI (the same month of last year=100) as the research object and select a total of 192 monthly data from January 2001 to December 2016 from the official website of the China National Bureau of Statistics; the data before September 2016 are used to train the model.

In this paper, we use four indicators including the MSE (defined earlier, see Eq. 11), root-mean-square error (RMSE, see Eq. 12), mean absolute error (MAE, see Eq. 13), and the mean absolute percentage error (MAPE, see Eq. 14) to evaluate the experimental results.

(a) Min: ARIMA (1, 0, 5)

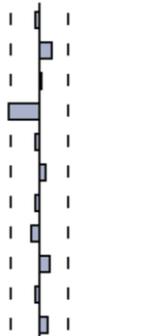

(b) Mean: ARIMA (1, 0, 2)

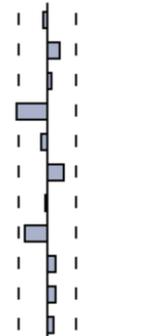

(c) Max: ARIMA (1, 0, 2)

**Figure 7:** White-noise test of ARIMA fitting residual

| Autocorrelation | Partial Correlation | | AC | PAC | Q-Stat | Prob |
|---|---|---|---|---|---|---|
| | | 1 | 0.002 | 0.002 | 0.0007 | 0.979 |
| | | 2 | 0.005 | 0.005 | 0.0048 | 0.998 |
| | | 3 | 0.032 | 0.032 | 0.1913 | 0.979 |
| | | 4 | 0.008 | 0.008 | 0.2031 | 0.995 |
| | | 5 | -0.073 | -0.074 | 1.2058 | 0.944 |
| | | 6 | 0.060 | 0.059 | 1.8794 | 0.930 |
| | | 7 | 0.075 | 0.075 | 2.9408 | 0.890 |
| | | 8 | -0.093 | -0.091 | 4.5814 | 0.801 |
| | | 9 | -0.123 | -0.129 | 7.4985 | 0.585 |
| | | 10 | -0.007 | -0.015 | 7.5087 | 0.677 |
| | | 11 | 0.012 | 0.031 | 7.5349 | 0.754 |

**Figure 8:** White-noise test of ARIMA (9,0,3) fitting residual

$$MSE = \frac{\sum_i^N (P_i - O_i)^2}{N} \tag{11}$$

$$RMSE = \sqrt{\frac{\sum_i^N (P_i - O_i)^2}{N}} \tag{12}$$

$$MAE = \frac{\sum_i^N |P_i - O_i|}{N} \tag{13}$$

$$MAPE = \frac{\sum_i^N \left|\frac{O_i - P_i}{O_i}\right|}{N} \tag{14}$$

On the whole, the experiment includes two parts. The first is to predict the monthly PPI sequence and its variation through the GA-SVR-ARIMA model based on fuzzy information granulation; the second is to verify the prediction ability of the specific method as more accurate and more suitable than other methods. The LIBSVM toolkit is used to test the SVM during the experiments.

## 4.2 Pre-classification of fuzzy information granulation

To study the seasonal variation of PPI and make accurate predictions, the research takes one quarter (3 months) as an information-granulation window to enlarge the time granularity for seeking the relative stable feature of the PPI monthly data by drawing on the experience of [Sun, Shao, Ji et al. (2014)]. In Matlab R2014a, we turn the original PPI sequence (Fig. 9) into three parameters—Min, Mean, and Max—through fuzzy granulation, all of which can reflect the seasonal variation trend of PPI (Fig. 10). To be specific, Min, Mean, and Max indicate the minimum value, the average value, and the maximum value of the single seasonal variation, respectively.

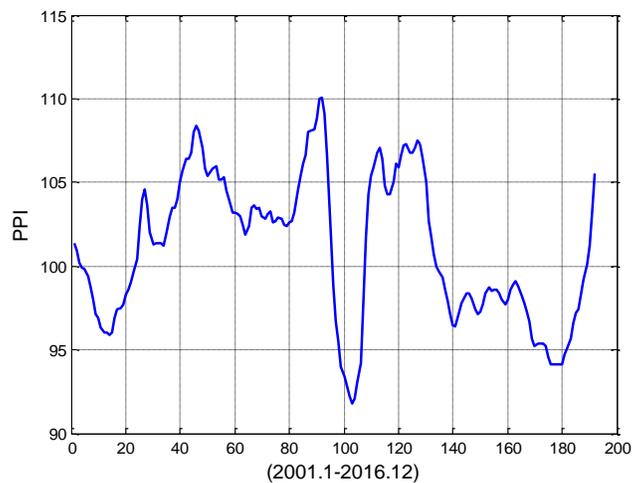

**Figure 9:** Original PPI sequence diagram

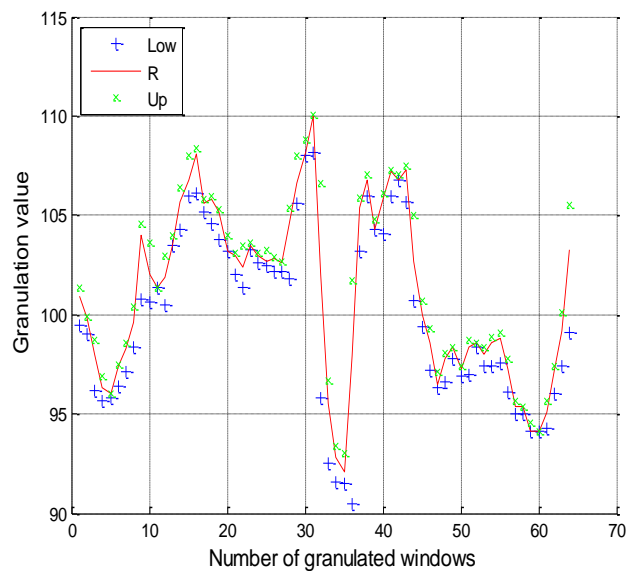

**Figure 10:** Fuzzy information-granulated sequence diagram

According to the results of fuzzy information granulation, the GA-SVR model is established to train and predict fuzzy information particles and windowed PPIs. Taking PPIs from January 2001 to September 2016 as a training set and PPIs from September 2016 to December 2016 as a test set, we regard one quarter (3 months) as an information-granulation window, and the input samples of the training set become a 63*3-type matrix after fuzzy information granulation.

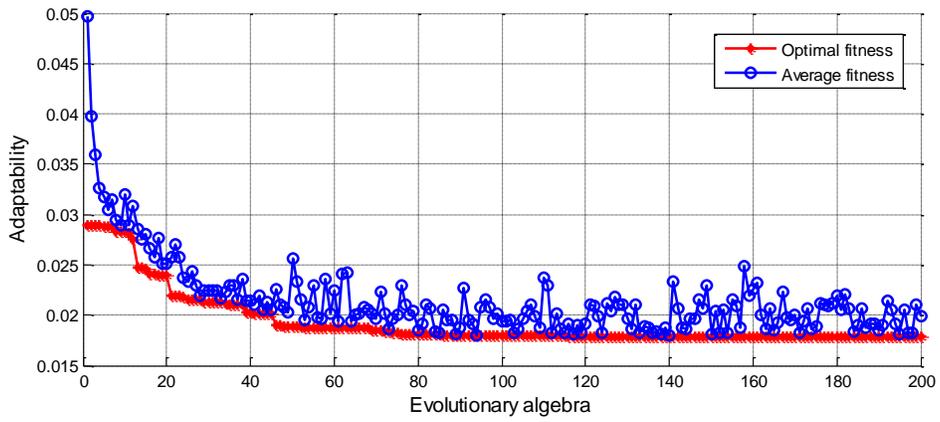

(a) Fitness curve of training set Min

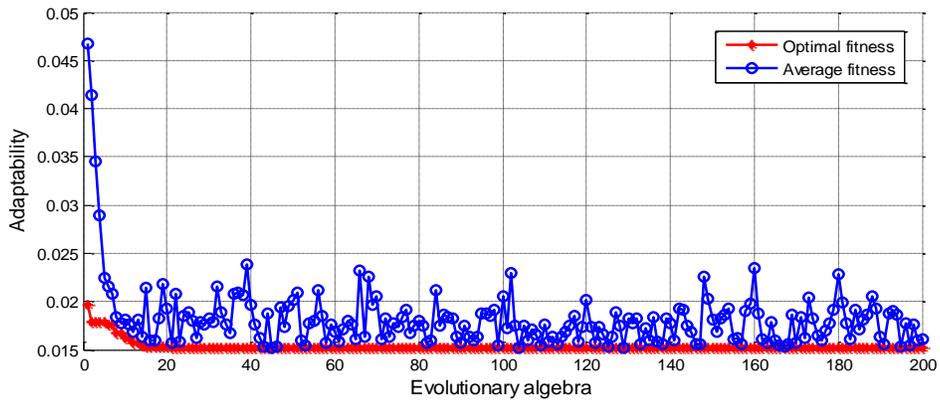

(b) Fitness curve of training set Mean

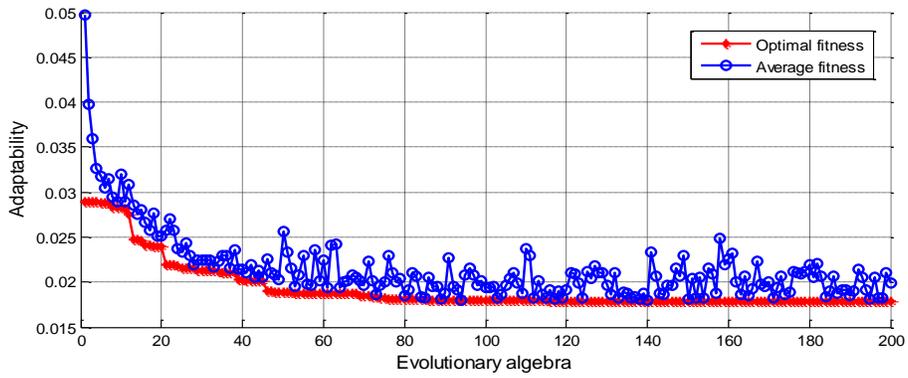

(c) Fitness curve of training set Max

**Figure 11:** MSE change chart on fitness curve of GA parameter optimization

*4.3 Parameter optimization based on GA*

Using Matlab R2014a, the GA is used to optimize the penalty parameter C, the insensitive loss function $\varepsilon$, and the parameter $\gamma$ of the RBF kernel function in the SVR model. We set the evolutionary algebra as 200 generations and the populations as 20 and choose the MSE as the fitness function for fivefold cross-validation. The process and the results of the parameter optimization based on the GA are shown in Fig. 11.

**Table 3:** Results of parameter optimization based on GA

| Information-granulation sequence | Best c | Best γ | Best $\varepsilon$ | MSE |
|---|---|---|---|---|
| Min | 0.093555 | 3.4514 | 0.17709 | 0.018627 |
| Mean | 0.23575 | 35.5006 | 0.11644 | 0.015116 |
| Max | 0.087261 | 2.8172 | 0.10179 | 0.019158 |

It is clear in Tab. 3 that the MSE values of the final optimization results of the three parameters are less than 0.02, which proves that the optimal parameters based on GA are more suitable.

*4.4 Prediction of GA-SVR model*

In view of the results of the optimization parameter based on the GA, the SVM model is established by using the results of penalty parameter C, insensitive loss function $\varepsilon$, and the parameter $\gamma$ of the RBF kernel function to predict the fuzzy information particles of the three parameters Min, Mean, and Max.

As is depicted in Fig. 12, the GA-SVR model fits the training set of three parameters well through information granulation and illustrates the relationship between nonlinear fluctuation and linear trend.

From Tab. 4, it is obvious that the overall fitting accuracy of the GA-SVR model is high. According to the order of MSE size, the prediction effect of the parameter Mean is the best, which is followed by the parameter Max, and the prediction effect of the parameter Min is relatively poor.

*4.5 ARIMA modeling of GA-SVR model residuals*

Using Eviews 8.0, we build the final GA-SVR-ARIMA hybrid model by using the ARIMA model for fitting the GA-SVR prediction residuals of the three parameters Min, Mean, and Max in the training set and predicting the testing set.

*4.5.1 Model fitting effect and predicting results*

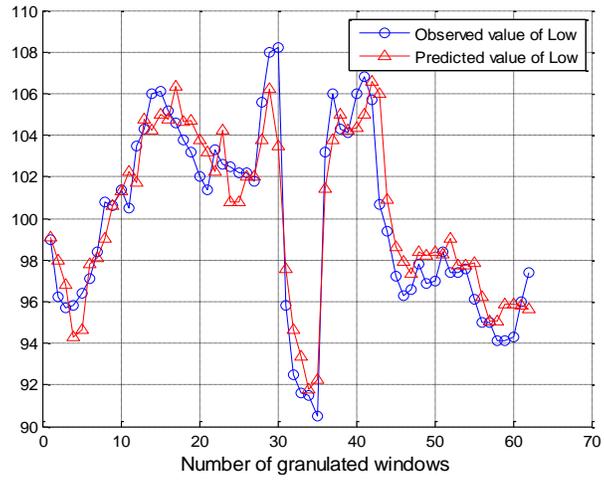
(a) Min parameter fitting

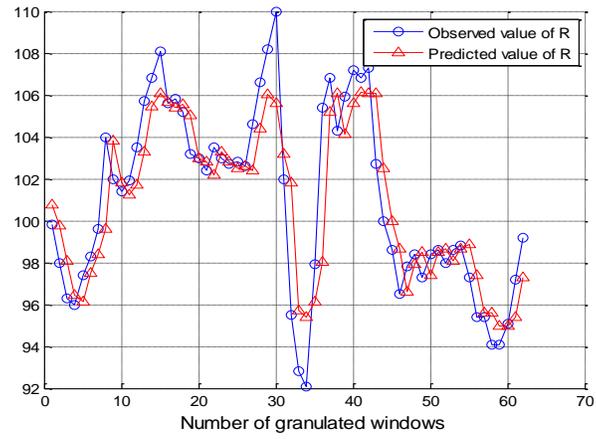
(b) Mean parameter fitting

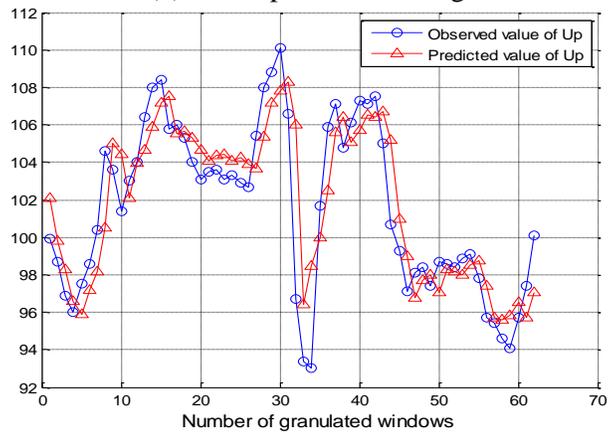
(c) Max parameter fitting
**Figure 12:** GA-SVR fitting results

**Table 4:** MSE of sequence prediction based on GA-SVR

| Test set of information-granulation parameters | Min | Mean | Max |
|---|---|---|---|
| MSE | 0.0173095 | 0.0140169 | 0.0188617 |

The proposed ARIMA model is used to fit the model, the results of which are shown in Fig. 13. Then, the ARIMA model is carried out for static prediction of the residual series of Min, Mean, and Max forecast by the GA-SVR. Finally, the variation range of the single season PPI is presented in Tab. 5.

To further assess the prediction effect of the fuzzy information-granulated GA-SVR model and GA-SVR-ARIMA hybrid model, the evaluation index of the test set is given in Tab. 6.

From Tab. 5 and Tab. 6, the five indexes of PPI predicted by the GA-SVR-ARIMA hybrid model established in this paper are obviously less than those of the GA-SVR single model, which indicates that the accuracy of the variation range predicted by the hybrid model is significantly higher than that of the single model.

Furthermore, these results illustrate that the PPI sequences have both non-linear and linear characteristics, and that the GA-SVR-ARIMA hybrid model can utilize the advantages of non-linear space prediction in the GA-SVR and linear space prediction in ARIMA.

Generally speaking, the PPI over 3 months maintains an upward trend, indicating that the prediction results could scientifically support the decision-making.

## 5 Comparative analysis of various methods

### 5.1 Comparative test of ARIMA model

In this paper, the PPIs from January 2001 to September 2016 are used as the training set, the input sample of which is a 189*1-type matrix. In addition, we take the PPIs from October 2016 to December 2016 as the test set, and then carry out the contrasting experiment in Eviews 8.0.

#### 5.1.1 Model fitting effect and prediction results

The ARIMA (9, 0, 3) model is adopted to fit the original training set, and the effect diagram is shown in Fig. 14.

Then, we use the established ARIMA (9, 0, 3) model for dynamic prediction, and the prediction values and results analysis are displayed in Tab. 7 and Tab. 8, respectively.

### 5.2 GRNN model comparison test

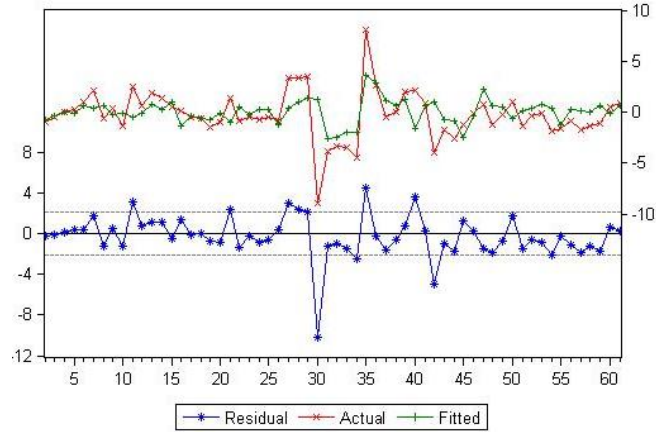

(a) Min model residual

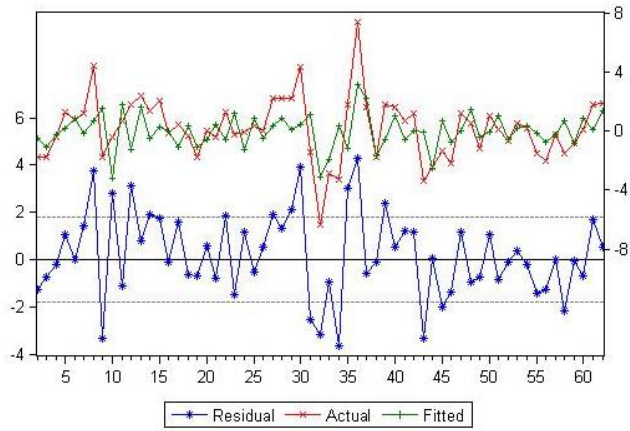

(b) Mean model residual

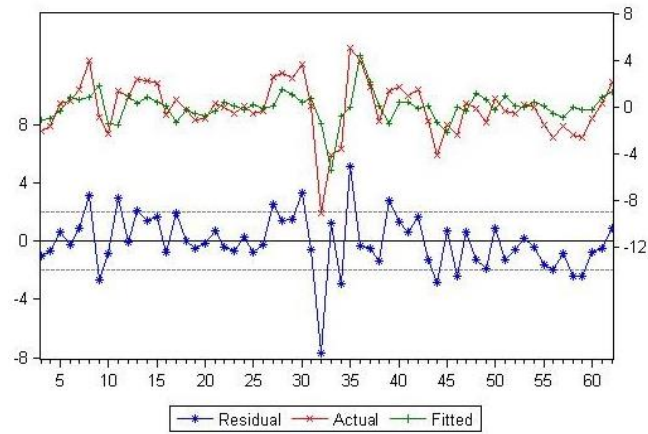

(c) Max model residual

**Figure 13:** ARIMA fitting effect of GA-SVR model residuals

**Table 5:** Summary of model prediction results

| Time | 2016.10 | 2016.11 | 2016.12 | Variation range of fuzzy particle |
|---|---|---|---|---|
| Value of observed PPI | 101.2 | 103.3 | 105.5 | {99.1, 103.3, 105.5} |
| GA-SVR prediction results | 101.2832 | 99.23 | 100.3076 | {99.1, 103.3, 105.5} |
| Results of GA-SVR residuals predicted by ARIMA | −0.0729 | 3.71 | 4.6758 | |
| Prediction results of GA-SVR-ARIMA | 101.2103 | 102.94 | 104.9834 | {99.1, 103.3, 105.5} |

**Table 6:** Analysis of prediction effect

| Statistics | MSE | RMSE | MAPE | MAE |
|---|---|---|---|---|
| GA-SVR model | 14.5109 | 3.8093 | 2.9813 | 3.1152 |
| GA-SVR-ARIMA hybrid model | 0.1322 | 0.3636 | 0.2828 | 0.2956 |

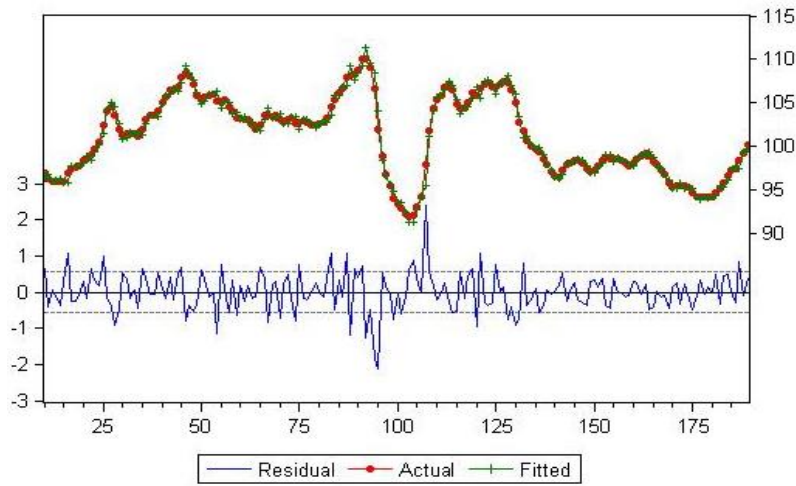

**Figure 14:** ARIMA fitting results of original PPI sequence

In the Matlab R2014a, the PPIs from January 2001 to September 2016 are selected as the training set, the input samples of which are a 189*1-type matrix, and the PPIs from October 2016 to December 2016 are used as the test set.

*5.2.1 Basic structure of GRNN model*

GRNN(General Regression Neural Network) was first proposed by [Specht(1993)] and its simple network structure based on GRNN in the Matlab Toolbox is shown in Fig. 15.

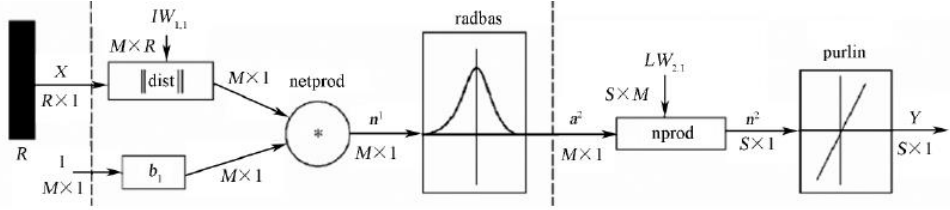

**Figure 15:** GRNN structure diagram

The input of the network is X, the total number of training samples is M, and the network output is Y. The first layer is the radial basis hidden layer, whose number of units is equal to the number of training samples M, and the weight function of the modified layer is dist, which contributes to calculating the distance between the network input and the weight of the first layer, $IW_{11}$:

$$\|dist\|_j = \sqrt{\sum_{i=1}^{R}(x_i - IW_{ji})^2}, (j=1,2,\cdots,M). \quad (15)$$

The network product function "netprod" forms the net input $n^1$ of the result of multiplying the element of the hidden layer threshold $b_1$ by the output element of "dist," and then passes it to the transfer function "radbas." Moreover, a Gaussian function is commonly used as the transfer function of the network:

$$a_j^1 = radbas(netprod(\|dist\|_j b_{1j})) = \exp\left|-\frac{(n_j^1)^2}{2\sigma_j^2}\right| = \exp\left|-\frac{(\|dist\|_j b_{1j})^2}{2\sigma_j^2}\right|$$

$$(j=1,2,\cdots,M) \quad (16)$$

where $\sigma_j$ is the smooth factor, which determines the shape of the basis function at the "*j*th" hidden layer, and the stationarity of the basis function is directly proportional to the value of $\sigma_j$. The second layer of the network is the linear output layer, the weight function of which is the normalized point weight product function (expressed by "nprod"), and the output at the previous layer and the dot product of the weight $IW_{21}$ at this layer are considered as the weight input, which can be directly passed to the transfer function purlin. The expression of the network output is

$$y_k = \sum_{j=1}^{m} IW_{kj} a_j^1 (k=1,2,\cdots,s). \quad (17)$$

*5.2.2 Method of determining smooth factor*

The smooth factor has great influence on the network prediction, so it is vital to select the appropriate smooth factor to improve the prediction accuracy. This experiment is optimized on the basis of the smooth factor determination proposed by [Sprecht(1993)] and finally adopts the one-dimensional optimization method to optimize the smooth factor. The optimization steps are summarized as follows:

1) Set an initial smooth factor $\sigma$.

2) Choose one sample from the training samples for testing and use the rest to construct the network.

3) Use the constructed network model to calculate the absolute value of error in the test sample.

4) Repeat steps (2) and (3) until all the training samples are used once for detection and calculate the mean-square error between all the predictive values to be estimated and the measured values of the samples:

$$MSE(\sigma) = \frac{1}{n}\sum_{i=1}^{n}\left[\hat{y}_t - y_t\right]^2 \qquad (18)$$

where $y_t$ is the training sample value and $\hat{y}_t$ is the predicted value of the network after training.

In this paper, we choose the initial smooth factor as 0.1 and add one unit quantity (0.1) each time, and then select the smooth factor as the optimal value when the average square error between the predictive value of the estimated point and the measured value of the sample is at the minimum. The average squared error between all the predictive values and the measured values is plotted in Fig.16.

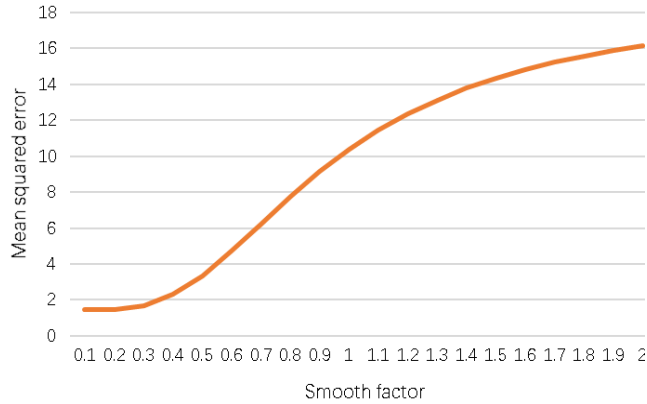

**Figure 16:** Determination of optimal smooth factor

Fig. 16 shows that when the smooth factor is 0.1, the error of both approximation and prediction is the smallest. As the smooth factor increases, the error keeps increasing as well. Thus, a GRNN network model with a smooth factor of 0.1 is selected for training and to facilitate comparison with the actual results (Fig. 17).

As Fig. 17 shows, the training results of GRNN network conform to the actual results, and the fitting results are good as well. To verify the intelligence and generalization ability, we employ the GRNN model to predict the remaining three monthly test samples. The results and analysis are shown in Tab. 7 and Tab. 8, respectively.

To further analyze the effect of various models on PPI sequence prediction, the evaluation indicators of the prediction set are given in Tab. 8.

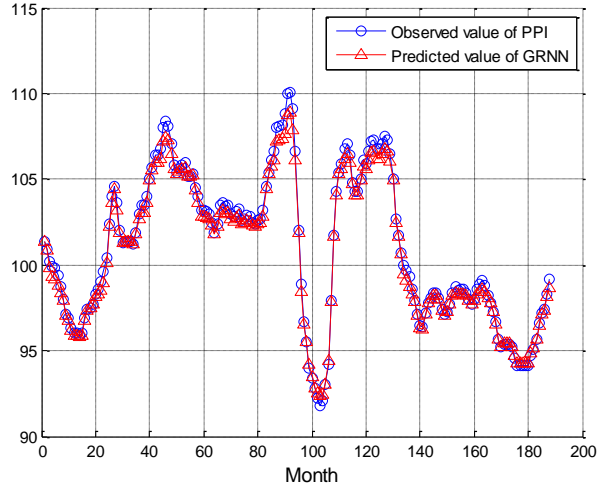

**Figure 17:** Fitting effect of GRNN training set

**Table 7:** Prediction results of test set

| Time | 2016.10 | 2016.11 | 2016.12 |
|---|---|---|---|
| Actual PPI value | 101.2 | 103.3 | 105.5 |
| ARIMA result | 100.971 | 104.954 | 106.864 |
| GRNN result | 99.992 | 101.517 | 103.212 |
| GA-SVR result | 101.2832 | 99.23 | 100.3076 |
| GA-SVR-ARIMA result | 101.2103 | 102.94 | 104.9834 |

**Table 8:** Analysis of prediction effect

| Statistic | MSE | RMSE | MAPE | MAE |
|---|---|---|---|---|
| ARIMA model | 1.5496 | 1.2448 | 1.0401 | 1.0823 |
| GRNN model | 3.2911 | 1.8141 | 1.6962 | 1.7597 |
| GA-SVR-ARIMA hybrid model | 0.1322 | 0.3636 | 0.2828 | 0.2956 |
| GA-SVR model | 14.51 | 3.809 | 2.981 | 3.115 |

It can be seen from Tab. 8 that the GA-ARIMA model with fuzzy information granulation gives the poorest prediction, and the ARIMA model performs better than the GRNN. In particular, the GA-SVR-ARIMA model based on fuzzy information granulation with the smallest prediction error has the best prediction effect. Ultimately, the results of the experimental comparison highlight that the GA-SVR-ARIMA hybrid model is more accurate in prediction.

After comparing several models, including GA-SVR, GA-SVR-ARIMA, ARIMA, and GRNN, through the statistical values of MSE, RMSE, MAPE, and MAE, we can rationally draw the conclusion that non-parametric and parametric hybrid models are more accurate and effective in prediction than a single model, which illustrates that the PPI sequences have both non-linear and linear characteristics.

We can therefore also draw the following conclusions based on the results calculated above:

1) The empirical study of PPI results in a relatively accurate prediction of the PPI trend, which can provide the basis for relevant decision-making around the world, which shows high practicability.

2) The predicted results of our "based on fuzzy information-granulated GA-SVR-ARIMA hybrid model" is the more accurate of all the relative research and traditional models, including non-parametric and parametric methods. At the same time, the variation and trend of prediction results agree well with reality, which, in turn, demonstrates the validity and practicability of the model.

3) The evidence that the statistical values, including the MSE, RMSE, MAPE, and MAE, of the hybrid model are smaller than other models indicates that complementarity of different models exists, including fuzzy information granulation, GA-SVR, and ARIMA.

## 6 Conclusions

The accurate forecast of the PPI is a major problem that has yet to be addressed in economic research. In the research presented herein, we first advanced a hybrid model scientifically based on fuzzy information granulation, the GA-SVR model, and the ARIMA model. What's more, the accurate results including the value and trend of PPI calculated using the proposed "based on fuzzy information-granulated GA-SVR-ARIMA hybrid model" play an extremely indispensable role in the processes of government management and other economic tasks that depend on the PPI. Our research has led to the development of a new exploration and measurement method based on our hybrid model that overcomes the problem of inaccurate PPI prediction. This hybrid model can comprehensively utilize the advantages of the component models, including fuzzy information granulation, GA-SVR, and ARIMA, making full use of the complementarity of different forecasting technologies. However, due to the short prediction period of this model and the neglect of other relevant variables, it remains to improve the model to make a more long-term forecast of PPI in light of various external factors, such as economic changes and price levels. Overall, our research contributes significantly to enhancing the accuracy and reliability of PPI forecasting by creative design and realization of the proposed algorithm. With the repaid development of cloud robotics [Liu et al. (2019)] and smart cities [Liu et al. (2017)], the method will be widely used. The method is especially useful for the development of the economic field.

**Acknowledgement**: This work was supported by the National Natural Science Foundation of China [ 61762033, 61702539]; The National Natural Science Foundation of Hainan [2018CXTD333, 617048]; Hainan University Doctor Start Fund Project [kyqd1328]; Hainan University Youth Fund Project [qnjj1444].